%% file: main.tex
\documentclass{JFM-FLM_Au}

\usepackage{tikz}
\usepackage{calc}
\usetikzlibrary{decorations.markings}
\usetikzlibrary{calc,arrows.meta,decorations.pathreplacing,positioning,shapes.geometric}
\usepackage[dvipsnames]{xcolor}
\usepackage{acronym}                      
\usepackage{layouts}
\usepackage[normalem]{ulem}   

\usepackage{titlesec}

\titleformat{\section}
  {\normalfont\Large\bfseries}{\thesection.}{1em}{}
  
\titleformat{\subsection}
  {\normalfont\large\bfseries}{\thesubsection.}{1em}{}

\titleformat{\subsubsection}
  {\normalfont\normalsize\bfseries}{\thesubsubsection.}{1em}{}

\acrodef{ROM}[ROM]{Reduced-Order Model}
\acrodefplural{ROM}[ROMs]{Reduced-Order Models}
\acrodef{SAM}[SAM]{Steady Actuation Manifold}
\acrodef{NRMSE}[NRMSE]{Normalized Root Mean Squared Error}
\acrodef{POD}[POD]{Proper Orthogonal Decomposition}

\lefttitle{A. Rodríguez-Asensio, G.Y. Cornejo Maceda, B.R. Noack, S. Discetti and A. Ianiro}
\righttitle{Feature-based manifold model of actuated wakes}

\title{Feature-based manifold model of actuated wakes}

\author{Alicia Rodríguez-Asensio\aff{1}, Guy Y. Cornejo Maceda\aff{1}, Bernd R. Noack\aff{2,1}, Stefano Discetti\aff{1} \and Andrea Ianiro\aff{1}}

\affiliation{\aff{1}Department of Aerospace Engineering, Universidad Carlos III de Madrid, Av. de la Universidad 30, Leganés, 28911, Madrid, Spain
\aff{2} College of Mechatronics and Control Engineering, Shenzhen University, Canghai campus, 518060, Shenzhen, PR China}

\corresau{Alicia Rodríguez-Asensio, \email{alicia.rodriguez.asensio@uc3m.es}}

\makeatletter
\def\ps@headings{%
  \def\@oddfoot{}%
  \def\@evenfoot{}%
  \def\@oddhead{\hfill\thepage}
  \def\@evenhead{\thepage\hfill}
}
\def\ps@firstpage{%
  \def\@oddhead{}
  \def\@evenhead{}%
  \def\@oddfoot{\hfill\thepage\hfill}
  \def\@evenfoot{\hfill\thepage\hfill}%
}
\makeatother
\begin{document}
\maketitle

\begin{abstract}
We propose a feature-based reduced-order model to predict the transient dynamics of bluff-body wakes under arbitrary time-varying actuation. 
Starting point is a control-oriented POD Galerkin modeling
which is challenged by incorporating time-varying actuations as a free input. Our model includes three key enablers.
First, POD modes are replaced by a more accurate feature-based manifold of same dimension.
Second, a state space is distilled from dynamic features which encapsulate time-varying coherent structures.
Third, this state space is augmented for the transient actuation response.
Thus, a simple analytical manifold dynamics is obtained.
The approach is applied to the fluidic pinball at $\Rey=30$, a canonical configuration of three identical circular cylinders arranged in an equilateral triangle and immersed in uniform flow under symmetric actuation. The model is validated against several representative actuation scenarios and accurately reproduces the transient dynamics without requiring unsteady training data, providing an interpretable, observable-based and control-oriented framework. The proposed description of actuated bluff-body flows is expected to be generalisable to other configurations.

\end{abstract}

\begin{keywords} 
 Low-Dimensional models, Wakes, Flow control
\end{keywords}


\section{Introduction}
\label{sec:introduction}
Reduced order modelling has reached a large maturity for uncontrolled and for steadily-controlled flows \citep{Brunton2015arm}. 
Yet, a control with enough authority should modify the flow so much that any data-trained reduced-order models  become inapplicable. 
In particular, transient control dynamics represent an additional challenge 
due to the appearance of new dynamics.
Under steady actuation, flow features may organise into a low-dimensional manifold that enables accurate and interpretable reduced-order modelling with actuation command as a tunable parameter \citep{marra2024actuation}. However, incorporating the control as a freely time-varying command has remained a challenge.
When the actuation varies in time, the flow state deviates from this \ac{SAM} and its evolution can no longer be described by steady-state mappings alone. This gap highlights a fundamental challenge: how to predict the transient evolution of low-dimensional flow features under time-varying control using only steady-state information. In this work, we address this problem by developing a feature-based manifold model that incorporates the distinct adaptation time scales of vortex-shedding dynamics and global wake adaptation represented through a delayed-actuation coordinate.

The existence of low-dimensional structure in bluff-body wakes rests on a well-established historical chain of evidence. \citet{landau1944problem} first introduced a phenomenological expression for the amplitude of a disturbance near a Hopf bifurcation. \citet{stuart1958non} subsequently derived this expression from the Navier--Stokes equations through an energy-balance argument, identifying the underlying physical mechanism as a mean-field correction: fluctuation growth on the unstable steady solution generates Reynolds stresses that deform the mean flow toward a less unstable state, throttling further amplitude growth. Building on this mean-field perspective, \citet{noack2003hierarchy} showed that the transient and post-transient cylinder wake can be captured by a low-dimensional Galerkin model in which mean-flow deformation is encoded by an additional shift-mode pointing from the unstable steady solution to the time-averaged flow. The resulting three-state dynamical system retains the quadratic nonlinearity of the Navier--Stokes equations, captures the globally stable limit cycle, and admits a slow-manifold interpretation: transients collapse onto a paraboloid in the amplitude--shift-mode plane, and the elimination of this slow direction recovers Landau's amplitude equation.
This framework has subsequently been extended to actuated flows by \citet{tadmor2010mean} for the cylinder wake, and by \citet{luchtenburg2009generalized} and \citet{semaan2016reduced} for high-lift configurations, with the latter explicitly modelling transients between the unactuated and the high-frequency-actuated attractors through a generalised mean-field model with multiple shift modes. 

Building on this perspective, recent data-driven approaches, such as sparse identification of nonlinear dynamics \citep{brunton2016discovering}, variational autoencoders \citep{wang2026information} and spectral submanifolds \citep{cenedese2022data}, have shown that highly nonlinear dynamics can be embedded in low-dimensional manifolds that capture essential flow physics. Notably, sparse identification has recovered the three-state mean-field model of \citet{noack2003hierarchy} directly from cylinder-wake snapshots \citep{brunton2016discovering}, confirming the structural relevance of the Stuart--Landau--Noack framework in purely data-driven settings. 
In particular, \citet{marra2024actuation} demonstrated that, for flows under steady actuation, the global wake state can be described by a low-dimensional manifold, the \ac{SAM}, embedded in a feature space. While effective when the actuation remains constant, such manifold descriptions do not account for transient deviations induced by time-dependent actuation and their robustness remain unclear \citep{haller2023nonlinear}.
When the actuation varies in time, the wake cannot instantaneously follow the \ac{SAM} due to intrinsic adaptation processes associated with vortex-shedding dynamics and mean-flow reorganisation. Predicting these transient deviations from only steady-state information remains an open challenge.

In this work, we investigate this problem using the fluidic pinball, a canonical configuration of three rotating cylinders exhibiting a wide portfolio of controlled-wake states in both laminar \citep{deng2020low,maceda2021stabilization} and turbulent \citep{rodriguez2026turbulent} regimes.
We consider symmetric actuation, either boat tailing (rear cylinders rotating inward, narrowing the wake and reducing drag) or base bleeding (rotating outward, enhancing the gap jet), that leads to wake stabilisation or destabilisation. The closest precedent is the generalised mean-field model of \citet{semaan2016reduced}, who captured transient actuation responses using a lifted \ac{POD} basis that explicitly accounts for the shift between unactuated and actuated attractors.
By contrast, the present work differs in three key aspects: we operate directly on observable force coefficients rather than \ac{POD} amplitudes; we exploit the continuous parametrisation of the SAM by the actuation parameter rather than a discrete set of attractors; and the model is driven solely by steady-actuation data, requiring no unsteady snapshots for calibration.
For the lift response, the formulation extends the three-state structure of \citet{noack2003hierarchy} through a delayed actuation coordinate $\tilde p$, which acts as a delayed actuation coordinate that encodes the global adaptation of the controlled wake to actuation changes; while the drag coefficient is governed by a first-order measurement equation. The dynamical state is therefore the oscillation amplitude, phase and the delayed actuation coordinate, while the drag coefficient is an output variable. 

This paper is organised as follows. The methodology is presented in \S\ref{sec:methodology}, including the flow configuration and model formulation. Results are discussed in \S\ref{sec:results}. Conclusions are offered in \S\ref{sec:conclusions}.

\section{Methodology}
\label{sec:methodology}

\subsection{Flow configuration}
\label{subsec:flowconfig}
The considered configuration is the fluidic pinball, consisting of a two-dimensional incompressible flow past a cluster of three identical circular cylinders of diameter $D$. They are arranged at the vertices of an equilateral triangle of side length $1.5D$ and pointing upstream. The Reynolds number is $Re=U_\infty D/\nu=30$, where $U_\infty$ is the inflow velocity and $\nu$ the kinematic viscosity. All quantities are expressed in convective units, based on $D$ and $U_\infty$. The simulations are performed with an in-house finite-element solver \citep{deng2020low} on an unstructured mesh of 15258 triangles and 30826 nodes, bounded in $[-20 ,80]\times[-30,30]$, with snapshots recorded every 0.1 convective units. Aerodynamic forces are non-dimensionalised using $\frac{1}{2}\rho D U_\infty^2$, where $\rho$ is the density of the fluid. 

Actuation is obtained by independently rotating the cylinder with tangential velocities $\mathbf{b}=(b_1,b_2,b_3)$. In this work, only symmetric actuations are considered, i.e., $b_1=0,\ b_2=-b_3$, expressed by the actuation (boat-tailing) parameter $p=(b_3-b_2)/2$. Positive values of $p$ correspond to boat tailing, which vectors the flow towards the centre and narrows the wake; while negative values induce base bleeding, which enhances the gap jet and prevents the development of a vortex near the cylinders \citep{maceda2021stabilization}.

The dataset comprises both statistically steady and transient simulations. For steady actuation, the actuation parameter is varied in the range $-1.5\leq p\leq1.5$, with steps in $p$ of 0.1. Each simulation is run for 1000 convective time units; the last 100 units are taken to characterise the corresponding limit cycle. The vortex-shedding amplitude becomes fully suppressed near $p\approx0.8$, thus the $p$ range is spanned with a finer resolution of $0.02$ in the range $0.7\leq p\leq 0.8$ to resolve the transition between regimes with and without vortex shedding. 
These steady-state responses define the steady actuation manifold, i.e.\ the locus of statistically stationary actuated states in the feature space introduced in \S\ref{subsec:model}, which maps the control parameter to the corresponding limit-cycle features. The full manifold is reconstructed under the assumption of smooth variations with respect to $p$, allowing intermediate states to be approximated via linear interpolation between neighbouring simulations.

A set of simulations is performed with time-varying actuation $p(t)$ to validate the model. They include ramp, periodic and smooth random variations between operating points, with normalised rates $0.1\leq \left|\frac{\mathrm{d}p}{\mathrm{d}t}\right|\leq1$, scaled by the unactuated vortex-shedding period. Representative trajectories are chosen to span both boat tailing and base bleeding regimes, including transitions from non-actuated flow ($p=0$) to strongly actuated states, as well as oscillatory forcing between distinct limit cycles.

\subsection{Feature-based manifold model}
\label{subsec:model}

The flow state is represented in a low-dimensional feature space composed of the lift and drag coefficients. The oscillatory dynamics are described in a phase-space representation using the lift coefficient $C_l(t)$ together with a delayed coordinate $C_l(t-\tau)$, where $\tau=T(p)/4$ corresponds to one quarter of the vortex-shedding period $T$ associated with the actuation level $p$. Within this representation, the oscillation amplitude and phase are defined as $A(t)=~\sqrt{C_l(t)^2+C_l(t-\tau)^2}$ and $\phi(t) =\tan^{-1}\bigl(C_l(t-\tau)/C_l(t)\bigr)$, respectively.

\begin{figure}[t]
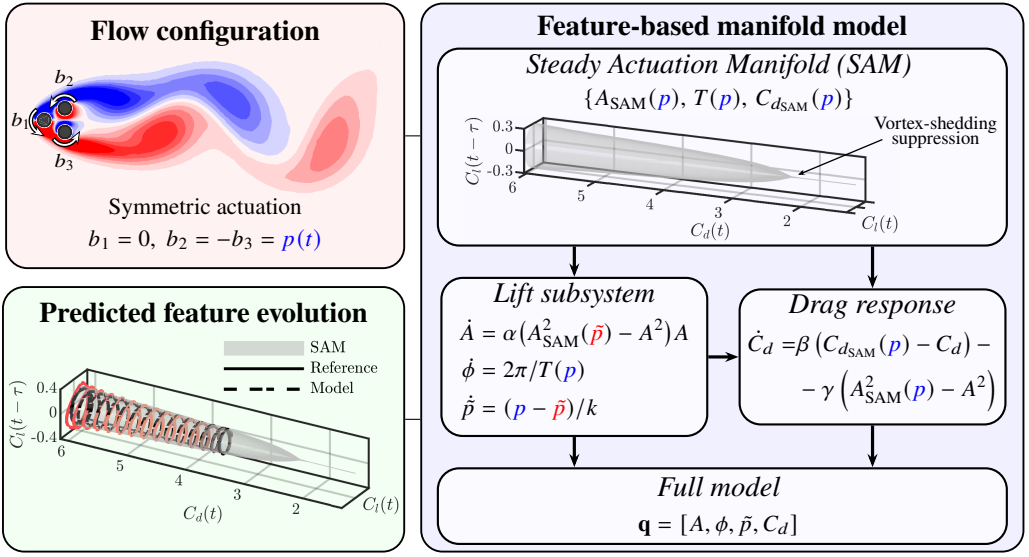

    \centering
    
    \include{Fig1_Methodology}
    \caption{Schematic of the feature-based manifold model for the transient wake dynamics of the actuated fluidic pinball. (\textit{Top left}) Flow configuration: the three-cylinder cluster at
    $Re=30$ under symmetric actuation $p(t)$. (\textit{Right}) Feature-based manifold model, consisting of three components: the steady actuation manifold, which provides the $p$-dependent functions; the lift subsystem, governing the evolution of the dynamical state $(A,\phi,\tilde p)$; and the drag response, a first-order measurement equation for $C_d$.  (\textit{Bottom left}) Predicted feature evolution: reference (solid) and modelled (dashed) transient trajectories shown together with the \ac{SAM} (gray) in the feature space $(C_l(t),\,C_l(t-\tau),\,C_d(t))$.}
    \label{fig:model_fig}
\end{figure}

Rotation of the rear cylinders deforms the mean wake organisation. Under time-varying actuation, two distinct responses emerge, which can be modelled separately. First, the vortex-shedding amplitude evolves on a slow time scale associated with the gradual reorganisation of the global wake. Second, the drag responds almost instantaneously to changes in actuation, but its final relaxation toward the limit-cycle value weakly depends on the vortex-shedding amplitude. Both responses are anchored to the \ac{SAM}: the saturated amplitude $A_\mathrm{SAM}(p)$, drag coefficient $C_{d_\mathrm{SAM}}(p)$ and shedding period $T(p)$ become smooth functions of $p$, all directly extracted from the steady simulations of \S\ref{subsec:flowconfig}. A schematic overview of the modelling framework is given in figure~\ref{fig:model_fig}. 

The starting point for the lift subsystem is the minimal Galerkin system derived by \citet{noack2003hierarchy}, which encodes the slow-manifold dynamics of the amplitude--shift-mode pair in the spirit of \citet{stuart1958non}:
\begin{align}
\frac{\mathrm{d}A}{\mathrm{d}t} &= (\mu - w)\,A, \\
\frac{\mathrm{d}\phi}{\mathrm{d}t} &= \omega_0, \\
\frac{\mathrm{d}w}{\mathrm{d}t} &= -w + A^2,
\end{align}
where $\mu>0$ is the linear growth rate of the most unstable eigenmode of the steady solution,  $\omega_0$ its angular frequency, and $w$ the shift-mode amplitude (the mean-field correction), which directly affects the oscillation amplitude and originates from the Reynolds-stress feedback of the fluctuating wake. The unstable steady solution corresponds to the fixed point $A=w=0$; whereas the saturated limit cycle is reached when $w=\mu$ and $A^2=\mu$.

In the present actuated framework, the role of $w$ is played by the delayed actuation coordinate $\tilde p$, but through a mechanism different from that of \cite{noack2003hierarchy}. While $w$ arises from internal Reynolds-stress feedback, $\tilde p$ encodes the delayed adaptation of the global wake to changes in imposed forcing. When the actuation $p$ changes, the wake reorganises gradually over several shedding periods, and $\tilde p$ tracks this process and determines the \ac{SAM}-based amplitude target toward which the oscillation amplitude relaxes. Adapting the Stuart--Landau--Noack framework to the lift oscillations of the actuated fluidic pinball gives the following dimensionless state equations:
\begin{align}
\frac{\mathrm{d}A}{\mathrm{d}t} &= \alpha\bigl(A_{\rm SAM}^2(\tilde p) - A^2\bigr) A, \label{eq:A}\\
\frac{\mathrm{d}\phi}{\mathrm{d}t} &= \frac{2\pi}{T(p)}, \label{eq:phi}\\
\frac{\mathrm{d}\tilde p}{\mathrm{d}t} &= \frac{p-\tilde p}{k}, \label{eq:ptilde}
\end{align}
The amplitude relaxes toward $A_\mathrm{SAM}(\tilde p)$, the \ac{SAM} target
at the current delayed actuation state. In contrast, the phase is driven by $T(p)$, at the instantaneous actuation rather than $\tilde p$, since the oscillation frequency responds to the immediate near-wake geometry. This asymmetry is consistent with the observation that frequency and amplitude respond on different time scales in actuated oscillator flows \citep{sipp2007global}. The dimensionless parameter $k$ reflects the convective persistence of the vortex street: it takes approximately four shedding periods for a newly shed vortex to be displaced far enough downstream that subsequent amplitude growth is governed by the current actuation rather than the shedding history, thus $k$ is taken equal to $4T(p)$. Here $k$ is evaluated at the instantaneous $p(t)$ via the \ac{SAM}, so it varies mildly during a transient.
The relaxation coefficient $\alpha$ controls the local amplitude rate toward the \ac{SAM} target. In practice, this parameter is weakly influential throughout most
of the \ac{SAM}, except near the vortex-shedding suppression threshold ($p\approx0.8$), where the topology of the manifold changes and sufficiently large values are required to ensure rapid collapse to the fixed point. For simplicity, we set $\alpha=k$, linking amplitude relaxation and wake adaptation through a single dimensionless characteristic scale, while noting that $k$ governs evolution of the delayed actuation coordinate and $\alpha$ determines how rapidly the amplitude adjusts once the delayed state has been established. 

On the other hand, the drag coefficient is modelled as a first-order output equation driven by the instantaneous actuation, consistent with its fast response relative to the amplitude dynamics and further adjusted with vortex shedding amplitude:
\begin{equation}
    \frac{\mathrm{d}C_d}{\mathrm{d}t} = \beta\left(C_{d_{\rm SAM}}(p) - C_d\right) - \gamma\left(A_{\rm SAM}^2(p) - A^2\right). \label{eq:Cd}
\end{equation}
In this formulation, the first term drives toward its \ac{SAM} target at the current actuation, while the second term provides a weak amplitude correction that ensures the drag coefficient converges to its asymptotic value only once the vortex-shedding amplitude itself has saturated. The coefficients $\beta$ and $\gamma$ characterise the adaptation scales. In the present work, both coefficients are set to unity, since this choice already provides satisfactory agreement with the transient trajectories without requiring additional calibration using unsteady data.

\begin{figure}[t]
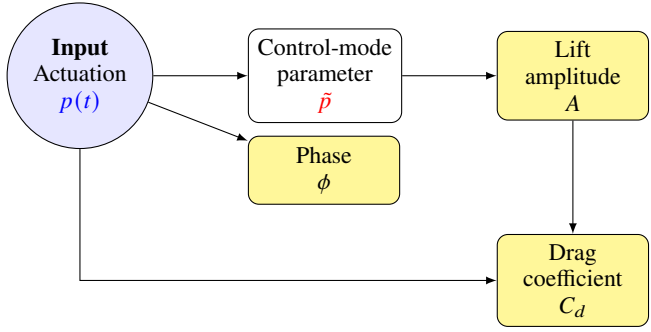

    \centering
    \include{Fig3_InfluenceDiagram}
    \caption{Influence diagram of the proposed feature-based manifold model showing the causal dependencies between the actuation input, delayed actuation coordinate, lift amplitude, phase and drag coefficient.}
    \label{fig:InfluenceDiagram}
\end{figure}

By construction, equations (\ref{eq:A})–(\ref{eq:Cd}) relax toward the SAM for constant $p$. Under time-varying forcing, the lag in $\tilde p$ produces transient amplitude deviations, whereas the drag responds rapidly to $p$ with a weak correction from the amplitude. The full model is $\mathbf{q}=[A,\phi,\tilde p,C_d]$, where the first three components constitute the lift subsystem and $C_d$ is an output variable governed by the response equation (\ref{eq:Cd}). The causal relationships between the variables are summarized in the influence diagram shown in figure~ \ref{fig:InfluenceDiagram}.

\section{Results}
\label{sec:results}
\begin{figure}[h!]
    \centering
    \includegraphics{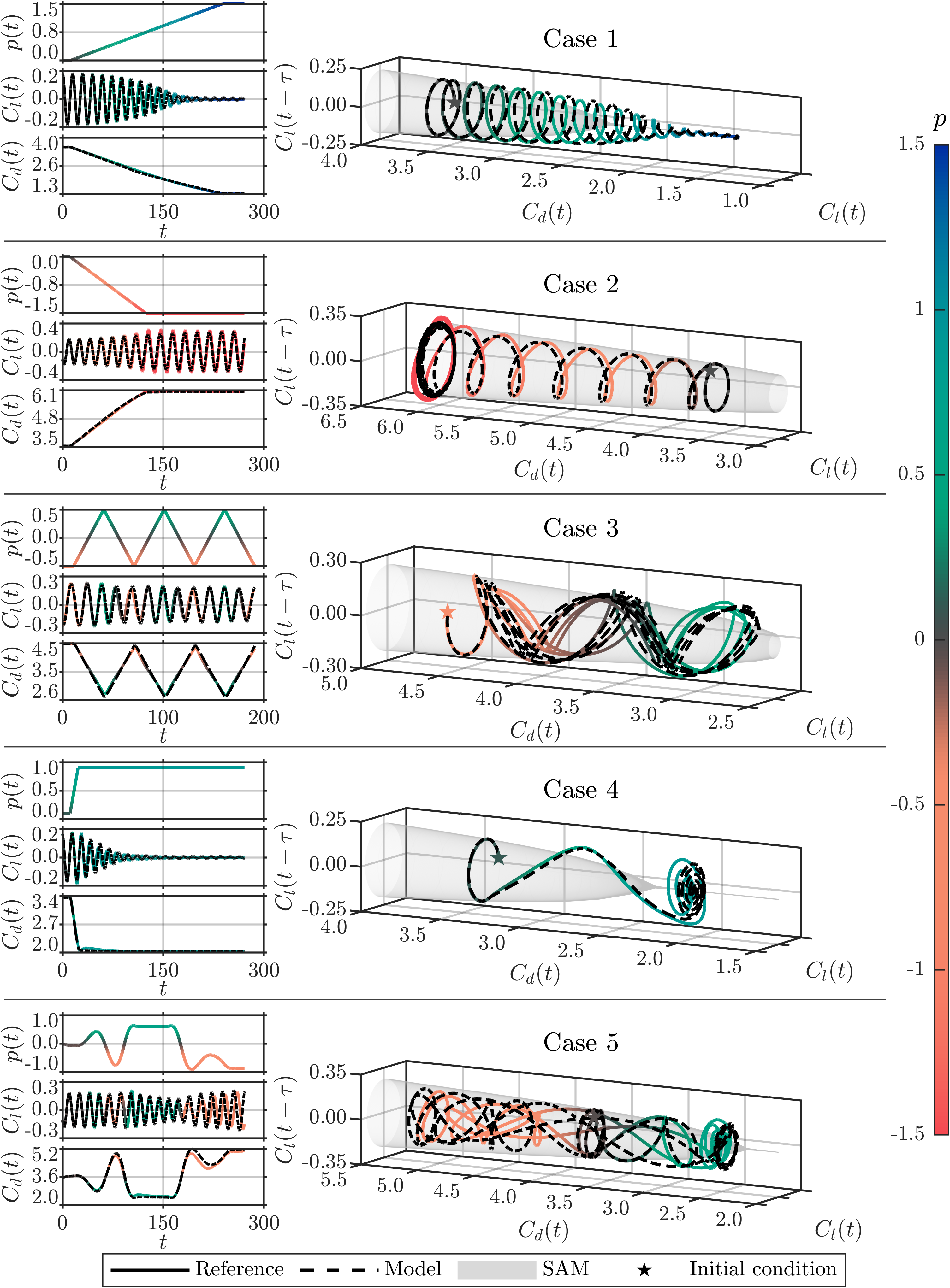}
    \caption{Comparison between reference and modelled trajectories for representative time-varying actuation cases of the fluidic pinball. For each case, the left column shows the temporal evolution of the actuation parameter $p(t)$, lift coefficient $C_l(t)$, and drag coefficient $C_d(t)$. The right column presents the corresponding trajectories in the feature space ($C_d(t)-C_l(t)-C_l(t-\tau)$). The \ac{SAM} is represented as a light-grey surface, the modelled trajectory by a dashed black line, and the reference trajectory by a coloured solid line. The star marker denotes the initial condition. The colour of the reference trajectory indicates the instantaneous actuation parameter, with warm colours corresponding to base bleeding, grey to the non-actuated configuration, and cool colours to boat tailing.}
    \label{fig:results}
\end{figure}

The feature-based manifold model is assessed against five representative trajectories of the fluidic pinball under time-varying symmetric actuation. Figure~\ref{fig:results} compares the modelled and reference trajectories in both temporal and manifold representation. Table~\ref{tab:results} summarises the quantitative results for the same trajectories, including the type of control law, the initial and final control, and the actuation rate $\frac{\mathrm{d}p}{\mathrm{d}t}$ which characterises the speed of the transient manoeuvrer. The model accuracy is assessed using the \ac{NRMSE} $E_x=\sqrt{\sum_{n=1}^{N}(x_n-\tilde{x}_n)^2/\sum_{n=1}^{N}x_n^2}$ computed for the lift amplitude $A$, drag coefficient $C_d$ and phase $\phi$. 
The trajectories are ordered by increasing complexity and actuation rate, from the slow linear ramp of Case~1 to the smooth random forcing of Case~5.

\begin{table}
    \centering
    \begin{tabular}{cccccccccc}
    Case & Control & Initial $p$ & Final $p$ & $\frac{\mathrm{d}p}{\mathrm{d}t}$  & $E_A(\%)$ & $E_{C_d}(\%)$ & $E_\phi(\%)$  \\
    \hline 
    1 & Linear & 0.0 & 1.5 & 0.1 & 6.61 & 1.30 & 2.87 \\
    2 & Linear & 0.0 & -1.5 & -0.2 & 11.5 & 0.635 & 0.382 \\
    3 & Periodic & -0.5 & 0.5 & 0.5 & 17.9 & 1.72 & 0.688  \\
    4 & Linear & 0.1 & 0.9 & 1.0 & 13.8 & 0.922 & 4.80 \\
    5 & Smooth random & 0.3 & 1.1 & - & 24.1 & 3.42 & 1.23
 
    \end{tabular}
    \caption{Quantitative assessment of the feature-based manifold model for representative transient actuation trajectories. Reported quantities include the type of control, initial and final actuation parameters, actuation rate (in units of the unactuated shedding period),  \ac{NRMSE} for lift amplitude, drag coefficient, and phase.
    }
    \label{tab:results}
\end{table}

Overall, the model accurately reproduces the transient wake evolution across a broad range of actuation scenarios: in all cases, predicted trajectories agree closely with the reference in both temporal and manifold representations. The agreement is particularly strong for the drag dynamics, with \ac{NRMSE} values consistently below $5\%$ for all cases. This is coherent with the structure of (\ref{eq:Cd}): because $C_{d_{\rm SAM}}(p)$ tracks the instantaneous actuation, the asymptotic drag target is always correct, and only the fast relaxation toward it contributes to the error. The model also captures the global organisation of the trajectories in feature space, including the gradual relaxation toward the steady actuation manifold. 
 
For slow and moderate transient manoeuvrers (Cases 1--3), the model reproduces both the lift amplitude and drag evolution with good accuracy. Case 1, corresponding to the slowest linear actuation, exhibits the smallest amplitude error, consistent with $|\mathrm{d}p/\mathrm{d}t|k \ll 1$: in this regime $\tilde p \approx p$ throughout the manoeuvrer and the trajectory stays close to the SAM. Although Cases 2 and 4 involve faster actuation, they yield even smaller drag errors than Case 1, because their trajectories do not approach the vortex-shedding-suppression region near $p\approx 0.8$. A noticeable deviation appears in this region, where vortex shedding becomes fully suppressed and the topology of the \ac{SAM} changes, in agreement with the stabilisation regime identified by \citet{maceda2021stabilization}. This is a known limitation of the proposed framework: the SAM was reconstructed under the assumption of smooth variation with respect to $p$ (\S\ref{subsec:flowconfig}), which breaks down at the Hopf bifurcation that suppresses vortex shedding. The model carries this assumption forward, so the local change in manifold topology is not transmitted to the dynamic equations. While the manifold geometry reflects this transition, the drag transient evolves more smoothly and thus does not immediately follow the local change in \ac{SAM} topology. Instead, the drag coefficient continues evolving approximately linearly toward the final steady state. For the fastest transient manoeuvrer (Case 4), the model remains qualitatively accurate but cannot reproduce small overshoots observed in the drag coefficient. These features fall outside the structure of (\ref{eq:Cd}), which is a first-order relaxation with a single time scale and cannot represent the higher-order response of the flow to rapid actuation; capturing them would require either a second-order drag equation or coupling $C_d$ to additional internal states. Similarly, the lift-amplitude error increases with actuation rate, as expected: as $|\mathrm{d}p/\mathrm{d}t|k$ grows, the control mode can no longer track the actuation closely, and the amplitude target $A_{\rm SAM}^2(\tilde p)$ in (\ref{eq:A}) systematically lags the saturated amplitude actually present in the reference simulation. This effect becomes particularly pronounced near the vortex-shedding suppression threshold, where the \ac{SAM} topology changes and the amplitude dynamics become sensitive to the coefficient $\alpha$.
 
The phase dynamics present a different behaviour. Equation~(\ref{eq:phi}) sets the model frequency to $2\pi/T(p)$ at the current actuation, but during transients, the dynamics temporarily deviate from the asymptotic \ac{SAM} frequency until the oscillation amplitude saturates. 
Since the phase error accumulates over time, longer trajectories naturally lead to larger deviations between the modelled and reference phase evolution. However, once vortex shedding is suppressed, the limit-cycle oscillation disappears, and the phase no longer evolves meaningfully. Because this transition is not explicitly embedded in the current formulation, the model cannot fully capture the phase dynamics beyond the suppression threshold. In a controlled-flow scenario, the instantaneous force measurements could be exploited to correct the accumulated phase delay.
 
The smooth random trajectory (Case 5) is the most demanding test and produces the largest errors for all features, except in phase. The actuation never remains constant long enough for $\tilde p$ to catch up with $p$ (the characteristic time between direction changes is comparable to $k$), so the amplitude target $A_{\rm SAM}^2(\tilde p)$ in (\ref{eq:A}) is itself in continuous motion and the system never enters a quasi-equilibrium regime. Nevertheless, the predicted trajectory still reproduces the main trends of the wake evolution and remains close to the reference manifold trajectory.

\section{Conclusions}
\label{sec:conclusions}
A feature-based manifold model has been developed for transient bluff-body wake dynamics under time-varying symmetric actuation, demonstrated on the fluidic pinball at $Re=30$. The model operates directly on observable lift and drag coefficients and is calibrated exclusively from steady-actuation data. The lift dynamics are governed by a three-state system extending the Stuart--Landau--Noack mean-field framework: the constant linear growth rate and frequency of the unactuated model are replaced by smooth functions of $p$ extracted from the \ac{SAM}, and the shift-mode amplitude is replaced by a delayed actuation coordinate $\tilde p$ that encodes the delayed global wake adaptation through a first-order relaxation. The drag response is modelled as a measurement equation driven by the instantaneous actuation, with a weak correction from the oscillation amplitude that ensures asymptotic convergence to the \ac{SAM}. The resulting dynamical system relaxes to the SAM for constant $p$ and predicts transient departures from it under time-varying $p$, without requiring unsteady training data. 

Given the initial conditions and the prescribed control law $p(t)$, the model accurately reproduces the deviations from the \ac{SAM} induced by time-dependent actuation for both wake stabilization and destabilization, including vortex-shedding suppression. Drag is reproduced with \ac{NRMSE} consistently below $5\%$ across all cases tested, a direct consequence of the design choice that the drag response depends on the instantaneous actuation rather than on $\tilde p$. The amplitude \ac{NRMSE} grows progressively with the actuation rate, becoming substantial as $|\mathrm{d}p/\mathrm{d}t|k$ becomes large, since the delayed actuation coordinate can no longer track the effective actuation. Two structural limitations of the formulation have been identified, both with minimal practical impact: the smooth-SAM assumption breaks down at the Hopf bifurcation that suppresses vortex shedding near $p\approx 0.8$, producing a localized and small deviation in the drag transient; and the first-order drag closure cannot represent the small overshoots observed in the fastest manoeuvrers, which would require either a second-order drag equation or coupling $C_d$ to additional internal states. In addition, the phase error accumulates progressively over time, leading to larger discrepancies for longer trajectories. Despite its limitations, the model successfully captures the dominant trends of the transient dynamics and associated mean-flow deformation. 

Overall, the present feature-based manifold model demonstrates that the stabilising and destabilising wake response to arbitrary time-varying actuation can be predicted directly in terms of the observable lift and drag coefficients, using exclusively steady-state information supplemented by physically meaningful adaptation time scales. As a result, the proposed framework provides an interpretable and computationally inexpensive reduced-order description of transient bluff-body wake dynamics, requiring no unsteady training data. Although demonstrated here using the laminar fluidic pinball, the methodology is expected to extend naturally to higher-Reynolds-number regimes, as analogous mean-field models based on the Stuart--Landau--Noack framework have been successfully generalized and applied to turbulent wake flows \citep{luchtenburg2009generalized,semaan2016reduced}. More broadly, the proposed model should be applicable to bluff-body flows exhibiting low-dimensional organisation under steady forcing, including wake stabilisation and drag-reduction problems relevant to aerodynamic applications. The presence of feature coordinates makes this model especially useful for implementation in a practical experimental framework in which lift and drag measurements could be exploited to take control decisions. In such frameworks, it is worth noting that, despite the presence of noise, the model accuracy could be further improved with periodic updates based on instantaneous measurements.


\begin{bmhead}[Acknowledgements]
During the preparation of this work, the authors used ChatGPT (OpenAI, GPT-5.5, accessed in 2026) in the writing process to improve the readability and language of the manuscript. After using this tool, the authors reviewed and edited the content as needed and assume full responsibility for the content of the published article.
\end{bmhead}

\begin{bmhead}[Funding.]
This publication is part of the project EXCALIBUR (Grant No. PID2022-138314NB-I00), funded by MICIU/AEI/10.13039/501100011033 and by ``ERDF/EU''. It is also part of the Grant PREP2022-000200 funded by MICIU/AEI/10.13039/501100011033 and by ESF+.
\end{bmhead}

\begin{bmhead}[Declaration of interests.]
The authors report no conflict of interest.
\end{bmhead}

\bibliographystyle{jfm}
\bibliography{jfm}



\end{document}

%% file: Fig1_Methodology.tex
\centering
\begin{tikzpicture}[
node distance=0.15cm,
input/.style={draw, fill=red!5, rounded corners=8pt, line width=0.25mm, minimum width=0.39\linewidth, minimum height=3.5cm, align=center},
model/.style={draw, fill=blue!6, rounded corners=8pt, line width=0.25mm, minimum width=0.59\linewidth, minimum height=7.25cm, align=center},
box/.style={draw, thick, fill=blue!2, rounded corners=8pt, minimum width=0.25\linewidth, minimum height=1cm, align=center},
boxin/.style={minimum width=0.1\linewidth, minimum height=3cm, align=center,draw=none},
arrow/.style={-{Latex[length=2.5mm]}, thick},
output/.style={draw, fill=green!5, rounded corners=8pt, line width=0.25mm, minimum width=0.39\linewidth, minimum height=3.5cm, align=center},
every node/.style={inner sep=2pt},
baseline,
]

\def\D{0.17}

\node[input, anchor=north west] (input) at (0,0) {
\textbf{Flow configuration}\\[-5pt]
\includegraphics{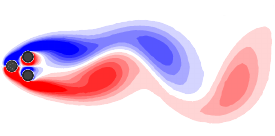}\\[-10pt]
\footnotesize Symmetric actuation\\[0pt] 
\footnotesize $b_1=0,\ b_2=-b_3={\color{blue} p(t)}$};

\node[model, anchor=north west] (model) 
at ([xshift=0.2cm]input.north east) {};
\node[font=\bfseries, anchor=north]
at ([yshift=-0.1cm]model.north) {
Feature-based manifold model};

\node[box, anchor=south] (fullmodel) at ([yshift=0.1cm]model.south) 
{ \begin{minipage}{0.54\linewidth}\centering
    \textit{Full model}\\[1pt]
    {\footnotesize $\mathbf{q}=[A,\phi,\tilde p,C_d]$
    }\\
  \end{minipage}
};

\node[box, anchor=south] (lift) at ([xshift=-0.145\linewidth,yshift=0.4cm]fullmodel.north) 
{ \begin{minipage}{0.25\linewidth}\centering
    \textit{Lift subsystem}\\[-12pt]
    {\footnotesize 
    \begin{align*}
    \dot A &= \alpha\bigl(A_{\rm SAM}^2({\color{red}\tilde p}) - A^2\bigr) A\\
    \dot \phi &= 2\pi/T({\color{blue} p})\\
    \dot{\tilde p} &= ({\color{blue} p}-{\color{red}\tilde p})/k 
    \end{align*}}\\
  \end{minipage}\\[2pt]
};

\node[box, anchor=west] (drag) at ([xshift=0.4cm]lift.east) 
{\begin{minipage}{0.25\linewidth}\centering
    \textit{Drag response}\\[-12pt]
    {\footnotesize    
    \begin{align*}
        \dot C_d = &\beta\left(C_{d_{\rm SAM}}({\color{blue} p}) - C_d\right) -\\ 
        &- \gamma\left(A_{\rm SAM}^2({\color{blue} p}) - A^2\right)
    \end{align*}}\\
\end{minipage}\\[2pt]
};

\node[box, anchor=south] (SAM) at ([xshift=0.145\linewidth,yshift=0.4cm]lift.north) 
{ \begin{minipage}{0.54\linewidth}\centering
    \textit{Steady Actuation Manifold (SAM)}\\[0pt]
    {\footnotesize $\{A_{\mathrm{SAM}}({\color{blue} p}),\,T({\color{blue} p}),\,C_{d_\mathrm{SAM}}({\color{blue} p})\}$}\\[1pt]
    \hspace{-1cm}\includegraphics{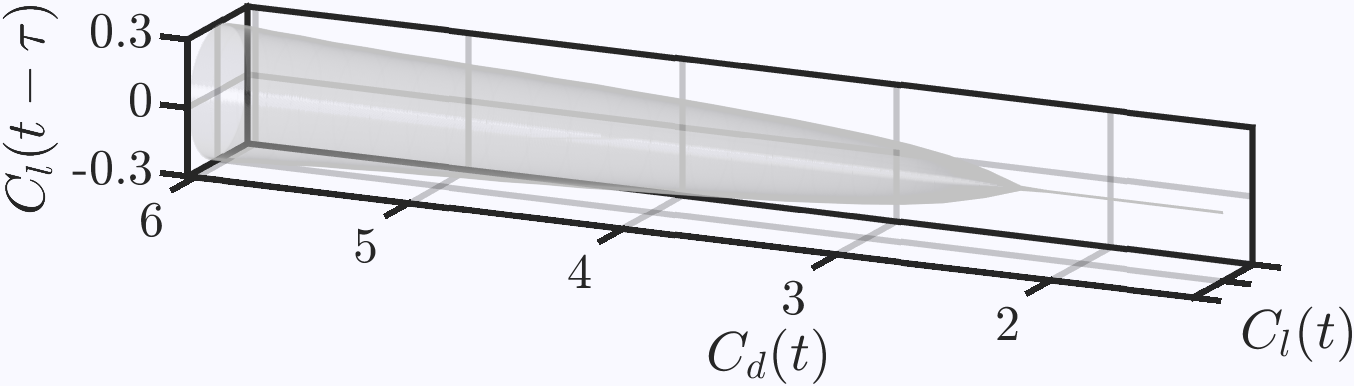}
  \end{minipage}
};

\node[font=\scriptsize, anchor=east] (vortex)
at ([xshift=-0cm,yshift=0.2cm]SAM.east) {
\begin{minipage}{0.13\linewidth}\centering
Vortex-shedding\\[-2.5pt]
suppression
\end{minipage}};
\coordinate (VortexSuppression) at ([xshift=1cm,yshift=0.95cm]SAM.south);
\draw[-{Stealth[scale=0.5]}, line width=0.5pt,black] ([xshift=0.25cm,yshift=-0.1cm]vortex.west) -- (VortexSuppression);

\coordinate (SAMLift) at (lift.center |- SAM.south);
\coordinate (SAMDrag) at (drag.center |- SAM.south);
\coordinate (VectorLift) at (lift.center |- fullmodel.north);
\coordinate (VectorDrag) at (drag.center |- fullmodel.north);
\draw[-{Stealth[scale=0.6]}, line width=1.2pt,black] (SAMLift) -- (lift.north);
\draw[-{Stealth[scale=0.6]}, line width=1.2pt,black] (SAMDrag) -- (drag.north);
\draw[-{Stealth[scale=0.6]}, line width=1.2pt,black] (lift.east) -- (drag.west);
\draw[-{Stealth[scale=0.6]}, line width=1.2pt,black] (lift.south) -- (VectorLift);
\draw[-{Stealth[scale=0.6]}, line width=1.2pt,black] (drag.south) -- (VectorDrag);

\coordinate (aC1) at ([xshift=0.51cm,yshift=0.2cm]input.west);
\coordinate (aC2) at ([xshift=0.785cm,yshift=0.365cm]input.west);
\coordinate (aC3) at ([xshift=0.785cm,yshift=0.05cm]input.west);

\path
  (aC1) ++(240:\D) coordinate (start)
  arc (240:120:\D) coordinate (end);

\draw[-{Stealth[scale=0.6]}, line width=1.3pt,black]
(start) -- ++(330:0.12);

\draw[draw, line width=1.75pt, black]
    (aC1) ++(120:\D) arc (120:240:\D);
\draw[line width=0.7pt,white]
    (aC1) ++(125:\D) arc (125:240:\D);
\draw[-{Stealth[scale=0.6]}, line width=0.6pt,white]
(start) -- ++(330:0.085);

\path
  (aC2) ++(150:\D) coordinate (start)
  arc (150:30:\D) coordinate (end);

\draw[-{Stealth[scale=0.6]}, line width=1.3pt,black]
(start) -- ++(240:0.12);

\draw[draw, line width=1.75pt, black]
    (aC2) ++(150:\D) arc (150:30:\D);
\draw[line width=0.7pt,white]
    (aC2) ++(150:\D) arc (150:35:\D);
\draw[-{Stealth[scale=0.6]}, line width=0.6pt,white]
(start) -- ++(240:0.085);

\path
  (aC3) ++(-150:\D) coordinate (start)
  arc (-150:-30:\D) coordinate (end);

\draw[-{Stealth[scale=0.6]}, line width=1.3pt,black]
(end) -- ++(-300:0.12);

\draw[draw, line width=1.75pt, black]
     (aC3) ++(-150:\D) arc (-150:-30:\D);
\draw[line width=0.7pt,white]
     (aC3) ++(-145:\D) arc (-145:-30:\D);
\draw[-{Stealth[scale=0.6]}, line width=0.6pt,white]
(end) -- ++(-300:0.085);

\node[font=\scriptsize]
  at ([xshift=-0.3cm,yshift=0cm]aC1)
  {$b_1$};

\node[font=\scriptsize]
  at ([yshift=0.4cm]aC2)
  {$b_2$};

\node[font=\scriptsize]
  at ([yshift=-0.4cm]aC3)
  {$b_3$};



\node[output, left=0.2cm of model.south west, anchor=south east] (output) {
\textbf{Predicted feature evolution} \\[6pt]
\includegraphics{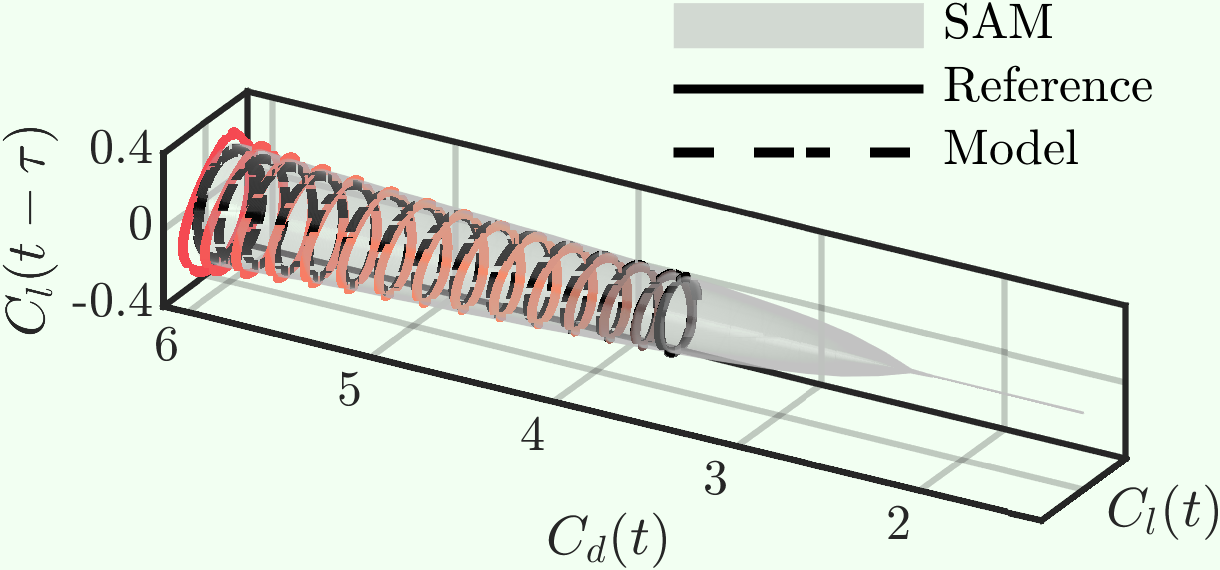}\\[-7pt]
};

\coordinate (modelinput) at (model.west |- input.center);
\coordinate (modeloutput) at (model.west |- output.center);
\draw (input.east) -- (modelinput);
\draw (modeloutput) -- (output.east);

\path[use as bounding box] (input.north west) rectangle (model.south east);
\end{tikzpicture}
\vspace{-0.75cm}

%% file: Fig3_InfluenceDiagram.tex
















\begin{tikzpicture}[
node distance=1.25cm,
>=latex,
every node/.style={font=\small},
block1/.style={draw, rounded corners, align=center,
minimum width=1cm, minimum height=0.9cm},
block2/.style={draw, rounded corners, fill=yellow!50,
align=center, minimum width=2cm, minimum height=0.9cm},
obs/.style={draw, circle, fill=blue!10,
align=center, minimum width=1cm}
]

\node[obs] (p) {\textbf{Input}\\
Actuation\\
${\color{blue} p(t)}$};

\node[block1, right=of p] (ptilde)
{Control-mode\\
parameter\\
${\color{red} \tilde p}$};

\node[block2, right=of p, yshift=-1.25
cm] (phase)
{Phase\\
${\phi}$};

\node[block2, right=of ptilde] (A)
{Lift\\ 
amplitude\\
$A$};

\node[block2, below=1.5cm of A] (Cd)
{Drag\\ 
coefficient\\
$C_d$};

\draw[->] (p) -- (ptilde);
\draw[->] (ptilde)  -- (A);

\draw[->] (p) -- (phase);

\draw[->] (A) -- (Cd);

\draw[->] (p) |-  (Cd);

\end{tikzpicture}
\vspace{-0.75cm}
